\documentclass[pra,twocolumn,showpacs,superscriptaddress,floatfix,nofootinbib]{revtex4}
\usepackage{color}
\usepackage{hyperref}  %%[hidelinks]
\usepackage{graphicx}
\usepackage{amsfonts}
\usepackage{footmisc}

\let\a=\alpha   \let\g=\gamma   
       \let\l=\lambda
\let\m=\mu    \let\n=\nu         \let\p=\pi    
\let\s=\sigma \let\t=\tau   \let\f=\varphi 
     
   \let\L=\Lambda 
         
\let\O=\Omega 

\font\tenmib=cmmib10\font\sevenmib=cmmib7\font\fivemib=cmmib5%
\def\BDpr {{\mbox{\boldmath$ \partial$}}}

\def\eqalign#1{\null\,\vcenter{\openup\jot
  \ialign{\strut\hfil$\displaystyle{##}$&$\displaystyle{{}##}$\hfil
      \crcr#1\crcr}}\,}

\def\AA{{\mathcal A}}
\def\EE{{\mathcal E}}\def\DD{{\mathcal D}}
\def\NN{{\mathcal N}}

\def\uu{{\V u}}\def\kk{{\V k}}\def\ff{{\V f}}
\def\T#1{{#1_{\kern-3pt\lower7pt\hbox{$\widetilde{}$}}\kern3pt}}

\def\ie{{\it i.e.\ }}
\def\dpr{{\partial}}
\def\defi{{\buildrel def\over=}}

\def\otto{\,{\kern-1.truept\leftarrow\kern-5.truept\to\kern-1.truept}\,}

\newdimen\xshift \newdimen\xwidth \newdimen\yshift \newdimen\ywidth

\def\ins#1#2#3{\vbox to0pt{\kern-#2pt\hbox{\kern#1pt #3}\vss}\nointerlineskip}

\def\eqfig#1#2#3#4#5{
\par\xwidth=#1pt \xshift=\hsize \advance\xshift
by-\xwidth \divide\xshift by 2
\yshift=#2pt \divide\yshift by 2
{\hglue\xshift \vbox to #2pt{\vfil
#3 \includegraphics{#4.eps}
}\hfill\raise\yshift\hbox{#5}}}

\def\V#1{{\bf #1}}
\def\lis#1{{\overline#1}}
\let\wh=\widehat

\def\tende#1{\,\vtop{\ialign{##\crcr\rightarrowfill\crcr
 \noalign{\kern-1pt\nointerlineskip} \hskip3.pt${\scriptstyle
   #1}$\hskip3.pt\crcr}}\,}
\def\eg{{\it e.g.\ }}

\def\0{\noindent}
\def\*{\vskip2mm}
\def\media#1{\langle #1 \rangle}
\def\Eq#1{\label{#1}}
\def\equ#1{(\ref{#1})}

\font\titolo=cmbx12%
\def\iniz{\setcounter{equation}{0}}
\def\be{\begin{equation}}\def\ee{\end{equation}}

\def\alertb#1{{\color{blue}#1}}

\newcounter{appendice}

\usepackage{fancyhdr}
\def\alert#1{{\color{ired}#1}}
\definecolor{iblue}{RGB}{65,105,225}
\definecolor{ired}{RGB}{220,20,60}
\definecolor{igreen}{RGB}{50,205,50}
\definecolor{ipurple}{RGB}{75,0,130}
\definecolor{iochre}{RGB}{218,165,32}
\definecolor{iteal}{RGB}{51,204,204}
\definecolor{imauve}{RGB}{204,51,153}

\iniz
\begin{document}

\alert{\centerline{\titolo Navier-Stokes equation: irreversibility,}
\centerline{\titolo turbulence and ensembles equivalence}}
\vskip1mm

\alertb{\centerline{\bf Giovanni Gallavotti}
  \centerline{\small INFN and U. Roma ``La Sapienza''} }
\centerline{\today}
%\maketitle %@

{\vskip3mm}
\noindent{\bf Abstract}: {\it The NS equation is considered (in $2\&3$
  dimensions) with a fixed forcing on large scale; the stationary states
  form a family of probability distributions on the fluid velocity fields
  depending on a parameter $R$ (Reynolds number). It is proposed that other
  equations could lead to -exactly- the same distributions via a mechanism
  closely analogous to the coincidence of the canonical and microcanonical
  averages of local observables in the statistical mechanics thermodynamic
  limit (proposed, here, to correspond to the limit in which the UV cut-off
  $N$, regularizing the equations, is removed to infinity).}
 
{\vskip3mm}
%\def\SEC{Introduction}
%\section{\SEC}
%\label{sec1}
%\iniz

%\input int

%{\bf This paper proposes an approach to the theory of the Navier-Stokes
%  equations simultaneously from the points of view of Dynamical Systems and
%  of Statistical Mechanics. The proposal is to look at viscosity as a
%  macroscopic manifestation of microscopic chaos and to obtain equivalent
%  descriptions of macroscopic systems. For stationary states equivalence is
%  seen in the same light as the equivalence of the equilibrium ensembles
%  ensembles in statistical mechanics: the role of the thermodynamic limit
%  is plaid here by the ultra-violet limit $N$ and full equivalence occurs
%  in the limit $N\to\infty$ in complete analogy with the equilibrium
%  ensembles equivalence in statistical mechanics in the thermodynamic limit
%  with volume (\ie infra-red cut-off) $V\to \infty$. The approach can be
%  used for $2$ and $3$ dimensional fluids: however evidence via example
%  simulations is, so far, provided in $2$ dimensions. The relation with the
%  scaling of Obuchov-Kolmogorov and the Ruelle-Lieb bounds is (very) briefly
%  commented. The aim of this note is to propose to a wider audience the
%  problem: because there are very efficient methods fully developed which
%  could be used for more severe tests in dimension $2$ and
%  for much needed tests in $3$ dimensions.}
%\*

The simplest geometry in which the Navier-Stokes equation can be studied is
a periodic geometry; the fluid is supposed incompressible and enclosed in
a periodic box, taken to be $\O=[0,2\p]^d,\ d=2,3$.  The velocity field
$\uu(\V x)$, together with its evolution equation, will be
represented via its Fourier's transform $\uu_\kk$ by:
\be \eqalign{
\uu(\V x)=&\sum_{\V0\ne\kk\in Z^d} \uu_\kk e^{i\kk\cdot\V x},
\kern2cm\V x\in\O\cr
\dot\uu_\kk\quad=&\ {\bf Q}(\uu)_\kk-\n\kk^2
\uu_\kk-\ff_\kk,\qquad \kk\cdot\uu_\kk=0\cr}
\Eq{1}\ee
where $\bf Q$ is quite simple in $d=2$ because, setting $\|\kk\|^2\defi$
$\sum_{\g=1}^d k_\g^2$ and $|\kk|\defi \max_{1\le \g\le d} |k_\g|$, 
 $\uu_\kk$ is essentially a (complex) scalar
$\uu_\kk=i\frac{\kk^\perp}{\|\kk\|} u_\kk$, $u_\kk=\lis u_{-\kk}\in C$,
while if $d=3$ $\uu_\kk=\lis \uu_{-\kk}$ is orthogonal to $\kk$; then ${\bf
  Q}(\uu)_\kk$ is
\be \eqalign{d=2&:\quad \sum_{\kk_1+\kk_2=\kk}
\frac{(\kk_1^\perp\cdot\kk_2) (\kk_2^2-\kk_1^2)}
     {2\|\kk_1\|\,\|\kk_2\|\,\|\kk\|}{u_{\kk_1} u_{\kk_2}}, \cr
     d=3&:\quad - \kern-3mm\sum_{\kk_1+\kk_2=\kk} i\,(\uu_{\kk_1}\cdot\kk_2)
     (\uu_{\kk_2}-\frac{\kk\cdot\uu_{\kk_2}}{\|\kk\|^2}\kk).
     \cr}\Eq{2}\ee
To avoid convergence and existence questions the $\kk,\kk_1,\kk_2$ and
the summations are restricted to waves with $1\le |\kk|,|\kk_1|,|\kk_2|\le N$ so
that the equations Eq.\equ{1},\equ{2} will actually define the NS
equations regularized with ``ultraviolet cut-off'' (``UV'') at
$N$. But {\it attention will be devoted to properties that admit a limit as
$N\to\infty$}.

Standing assumptions on the forcing $\ff$ will be, aside from
 $\kk\cdot\ff_\kk=0$, that $|\ff_\kk|\equiv0$
for $|\kk|>k_f$ with $k_f$ independent on the UV cut-off $N$, so that the
scale of $\ff$ will remain $k_f^{-1}$, \ie ``large'', as the UV cut-off
scale $N^{-1}\to0$, furthermore $\ff$ will be 
{\it fixed throughout the discussion} and supposed of size $1$, \eg
$\|\ff\|^2=\sum_\kk|\ff_\kk|^2=1$, $\|\ff |\kk|\|\le k_f$.

The regularized equations have smooth solutions for all initial data if
$\n>0$ and eventually satisfy $N$-independent {\it a priori}
bounds\footnote{\ie any initial data $\uu$ for time large enough satisfy a
  bound larger then the one called here {\it a priori} average.
  Also if the time average of $\DD_4(\uu)\defi \sum_\kk \kk^4|u_\kk|^2$
  exists and $d=2$ it is $\le k_f^4\n^{-2}$.} namely, if
$\min_{\kk\ne\V0} \|\kk\|=1$:
\be \eqalign{
&\EE(\uu)\defi\sum_\kk |\uu_\kk|^2\le
 \n^{-2},\qquad 
   d=2,3\cr
 &\DD(\uu)\defi\sum_{\kk} \kk^2|u_\kk|^2\le
  k_f^2\n^{-2}, \quad d=2\cr
}\Eq{3}\ee
bounding respectively energy $\EE(\uu)$, enstrophy $\DD(\uu)$.

Let $R\defi\n^{-1}$; then, for each $R$, the evolution will lead, as time
$t\to\infty$, a given intial datum $\uu$ randomly chosen (with respect to a
distribution with a smooth density) in $M_N$=space of the velocity
fields\footnote{$M_N$ dimension is $\NN=4N(N+1)$ if $d=2$, $\NN=O(8 N^3)$
%8N^3+12 N^2+6N$
if $d=3$.\label{dimension}} to a stationary state $\m^N_R$, which
depends on the UV cut-off.  The state $\m^N_R$ may be not unique and,
particularly if $d=3$, reaching it may require selecting a sequence
$t\to\infty$ of times.

Such stationary states form a set
$\EE^N_R$.  In regimes of strong turbulence (\ie $R$ large) it can be
expected that $\m^N_R$ is just unique and reached without attention to the
considered sequence of times $t\to\infty$.

The aim of this paper, pursuing the analogies between turbulence and
statistical mechanics, \cite{Ru989,Ru012,Ru014}, is to study whether
properties of the stationary states in $\EE^N_R$, which become {\it
  independent} of the UV cut-off $N$ as $N\to\infty$, are shared by
properties of stationay states of {\it other equations of motion} on the
same phase space $M_N$, \cite{SJ993,Ga997b}.

Here an {\it exact equivalence} is proposed between the equation obtained
by replacing $\n\kk^2 u_\kk$ by $\a(\uu) \kk^2 \uu_\kk$ with the multiplier
$\a(\uu)$ chosen so that $\DD(\uu)=\sum_\kk \kk^2|\uu_\kk|^2$, has a value
$En$ {\it exactly constant} in time; see Eq.\equ{8} below for an explicit
expression of $\a(\uu)$, \cite{Ga018}.

  The stationary states of the two equations form two collections
  $\EE^{i,N},\EE^{r,N}$ (labels $i,r$ for irreversible,reversible) whose
  distributions $\m^{i,N}_R,\m^{r,N}_{En}$ are parameterized respectively
  by the Reynolds (or Grashof) number $R$ and by the initial data enstrophy
  $En$ (equal to the constant value of $\DD(\uu)$) and depend on the UV
  cut-off $N$. A pair of such states will be called {\it corresponding} if
  the parameters $R,En$ are such that
  \be 
  \m^{i,N}_R(\DD)=En
  \Eq{4}\ee
which will acquire, see below, the physical meaning that the work done per
unit time by the forcing is the same in corresponding stationary
distributions. Proposal:
  
\* \0{\it %Conjecture:
  Let $O$ be any observable localized\footnote{\ie
  $O$ is a function of the Fourier's components $\uu_\kk$ with $|\kk|<K$.}
    on waves $\kk, |\kk|<K$, and a pair of corresponding states $\m^{i,N}_R$
    and $\m^{r,N}_{En}$ for the two equations above, then:
  \be
  \eqalign{
    &\lim_{N\to\infty}\m^{i,N}_R(O)= \lim_{N\to\infty}\m^{r,N}_{En}(O)
    \qquad{\rm for\ all}\ K<\infty.\cr}
  \Eq{5}\ee}
%\def\SEC{Heuristic analysis}
%\section{\SEC}
%\label{sec2}
%\iniz
This is inspired by the existence in equilibrium statistical mechanics of
``equivalent ensembles'': where suitably corresponding stationary
distributions which are very different nevertheless attribute the same
average values to large classes of observables ``in the thermodynamic
limit''. For instance the microcanonical and the canonical distributions at
fixed energy density and, respectively, at fixed average kinetic energy
density can be considered stationary states for suitable Hamiltonian
equations in which the total energy $E$ is conserved or, respectively, the
average kinetic energy is conserved, \cite{Ru969,Ru014}.

In these cases it is possible to extablish a correspondence between
stationary distributions, also called ``states'', in
such a way that observables $O$ which are {\it local}, \ie which depend on
particles located in a region $\L$ of space, small compared to the total
volume $V$ containing the system, will have the same average value in
corresponding stationary states. {\it Exactly} the same in the limit in
which $V\to\infty$ with $\L$ remaining fixed: \ie in the {\it thermodynamic
  limit}.

The analogy proposed here is to consider the UV cut-off $N$ as
playing the role of the total volume in statistical mechanics, with the
role of the energy being played by the enstrophy $\DD(\uu)$ and the role of
the inverse temperature being plaid by the Reynolds number $R$. 

The correspondence between stationary states $\m^{i,N}_R$ of the
Eq.\equ{1} and states $\m^{r,N}_{En}$ of the equation with the
``variable viscosity'' $\a(\uu)$ replacing $\n$, as above, is established
by Eq.\equ{4} and $N\to\infty$ is analogous to the thermodynamic limit.

The intuition behind the proposal is that, at least for large $R$, in the
variable viscosity equation $\a(\uu)$ fluctuates strongly generating a
{\it homgeneization} phenomenon implying that the local observables, being
variable only at large scale, ``see'' $\a(\uu)$ as a constant equal to its
time average. This argument fails in the cases in which the stationary
states are ``laminar'', \ie non chaotic: yet it will be argued that the
equivalence holds even in such cases.
%: where it can even be sometimes proved
%rigorously.

Further motivation is obtained remarking that the NS equations follow, via
suitable scaling limits, from the intermolecular microscopic equations. The
latter are reversible while the Eq.\equ{1} are irreversible because the
time reversal $I\uu=-\uu$ does not anticommute with the evolution flow,
denoted $\uu\to S_t\uu$: \ie $S_t I\ne IS_{-t}$.

Since the time reversal symmetry is a fundamental one, it should be
possible to describe the fluid phenomena on their macroscopic scale also
via reversible equations: irreversibility is conceptually due to the
chaotic motions and only phenomenologically represented by the viscosity
coefficient. The equations defined above by replacing $\n$ by $\a(\uu)$, so
built that the enstrophy $\DD(\uu)$ is exactly constant, are reversible
(because $\a(-\uu)\equiv -\a(\uu)$) and provide a natural test of the idea.

Looking for equations equivalent to the NS equations on the local
observables the above choice is also somewhat privileged.

Because imposing that the averages of the observable $W(\uu)=\sum_\kk\lis
\ff_k\cdot\uu_\kk$ (\ie the power injected per unit time by the forcing,
which is a local observable by the standing assumption on $\ff$) be the
same in corresponding states, is implied by\footnote{This follows by
  mutiplying both sides of the equations by $\lis \uu_\kk$ and summing over
  $\kk$.\label{identity}} $\frac1R\m^{i,N}_R(D)=\m^{r,N}_{En}(\a)
En$. Hence the property $\m^{r,N}_{En}(\a)=\frac1R$, on which the
homogeneization interpretation is based, is the same as the
$\m^{i,N}_R(\DD)=En$ if the conjecture holds. Hence checking that:
\be \lim_{N\to\infty} R\m^{r,N}_{En}(\a)=1, \Eq{6}\ee
is a first check, with a simple physical interpretation,
of the proposed claims.

In statistical mechanics, however, there are many other ensembles
equivalent to the canonical or microcanonical. So the conjecture can
possibly be extended to other equations, in which other observables are
imposed to be exactly constant, by replacing the $\frac1R=\n$ in NS by a
suitable multiplier.

For instance $\a$ could be chosen so that the energy $\EE(\uu)=\sum_\kk
|\uu_\kk|^2$ is exactly constant and local observables have equal averages
in corresponding stationary states $\m^{i,N}_R,\wh \m^{r,N}_E$ of the
irreversible equation Eq.\equ{1} and, respectively, of the new equation:
with the correspondence established by the condition $\m^{i,N}_R(\EE)=E$,
analogous to Eq.\equ{4}.

Then corresponding states $\m^{i,N}_R,\wh \m^{r,N}_E$ would be
determined by $\m^{i,N}_R(\EE)=E$; and ``equality of averages of local
observables in corresponding distributions'' should imply that the averages
of the local observable $W(\uu)\defi\sum_\kk \ff_\kk\cdot\uu_\kk$, power
injected by the forcing, \ie $\frac1R \m^{i,N}_R(D)\equiv\m_R^{i,N}(W)$ and
$\wh\m^{r,N}_E(\a D)\equiv\wh\m^r_E(W)$\footnote{Obtained as in the above
  footnote\footref{identity}.} {\it are equal}, leading to the test:
\be\lim_{N\to\infty}\frac1R \m^{i,N}_R(\DD)=\lim_{N\to\infty}
\wh\m^{r,N}_E(\a \DD)\Eq{7}\ee
which has the same interpretation as Eq.\equ{6}, \ie the average energy
dissipated per unit time is the same in two equivalent evolutions.

Of course if this is correct also other equations can be imagined which are
equivalent to the NS equations, \cite{Ga997b}.

%\def\SEC{NS: dimension 2 and 3}
%\section{\SEC}
%\label{sec3}
%\iniz

Returning to the conjecture, consider the equations
$\dot\uu_\kk=Q_\kk(\uu)-\a \kk^2\uu_\kk -\ff_\kk$ with $\a=\frac1R$: these
are irreversible equations.
%as the time reversal $I\uu=-\uu$ does not
%anticommute with the evolution flow, denoted $\uu\to S_t\uu$: \ie $S_t I\ne
%IS_{-t}$.
If, instead, $\a(\uu)=\a_d(\uu)$ for $d=2,3$, respectively, is defined so
that the enstrophy $\DD(\uu)=\sum_\kk\kk^2|\uu_\kk|^2$ is constant, \ie (with
$\kk=\kk_1+\kk_2$) 
\be
\eqalign{
\a_2(\uu)=& \frac
    {\sum_\kk\kk^2 \lis \ff_\kk\cdot \uu_\kk}
    {\sum_\kk\kk^4|\uu_\kk|^2}
    \cr
\a_3(\uu)=&\a_2(\uu)-i
  \frac{\sum_{\kk_1,\kk_2} \kk^2(\uu_{\kk_1}\cdot\kk_2)
    (\uu_{\kk_2}\cdot\uu_{-\kk})}{\sum_\kk \kk^4|\uu_\kk|^2}
   \cr
}\Eq{8}\ee
and evolution is {\it reversible} %because $\a(\uu)$ is odd in $\uu$.

The equivalence condition $En=\m^{i,N}_R(\DD)$ yields, if $W(\uu)=\sum_\kk
\lis \ff_\kk\cdot\uu_\kk$, that the identities\footnote{Obtained as in the above
  footnote\footref{identity}.}  $\frac1R\m^{i,N}(\DD)\equiv\m^{i,N}_R(W)$ and
$\m^{r,N}_{En}(W)\equiv \m^{r,N}_{En}(\a) En$ should imply, as mentioned
before Eq.\equ{6}, equality of the averages of the injected average power
$W$ (as $W$ is a local observable). Hence for $\a=\a_i,i=2,3$ the
mentioned test of the equivalence in Eq.\equ{6} should hold reflecting the
heuristic idea of equivalence as a homogeneization phenomenon.

Eq.\equ{6} takes a particularly simple form if $d=2$:
\be\lim_{N\to\infty}\,R\,
\m^{r,N}_{En}\Big(\frac{\sum_\kk \kk^2\lis \ff_\kk\cdot\uu_\kk}{\sum_\kk
  \kk^4|\uu_\kk|^2}\Big)=1\Eq{9}\ee
while if $d=3$ it  takes the form:
\be \eqalign{
  \lim_{N\to\infty}&\,R\,
  \m^{r,N}_{En}\Big(\a_2(\uu)\cr
  &-i
  \frac{\sum_{\kk_1,\kk_2}\kk^2 (\uu_{\kk_1}\cdot\kk_2)
    (\uu_{\kk_2}\cdot\uu_{-\kk})}{\sum_\kk \kk^4|\uu_\kk|^2}\Big)=1\cr}
  \Eq{10}\ee
The above relations can be seen as {\it sum rules} which test quite
critically the OK41-scaling laws, \cite[Ch.5]{Ga002}: naive application of
the OK scaling in $d=2$ would give a $\log R$ divergence of the limit in
Eq.\equ{9}, while in $d=3$ the $\a_2$ contribution to Eq.\equ{10} would
tend to $0$ as $N\to\infty$ (as $R^{-\frac12}$).  Hence the first relation
is a sum rule that depends strongly on the details of the OK laws and their
corrections, while in $d=3$ the second contribution, after $\a_2$, involves
the whole inertial range and is more delicate.

A few simulations can be invoked to support the above ideas in
$d=2$; in all of them the initial data rare random fields with enstrophy
prefixed at its average computed in a preliminary run;
a first among them is illustrated below:

\eqfig{190}{125}{}{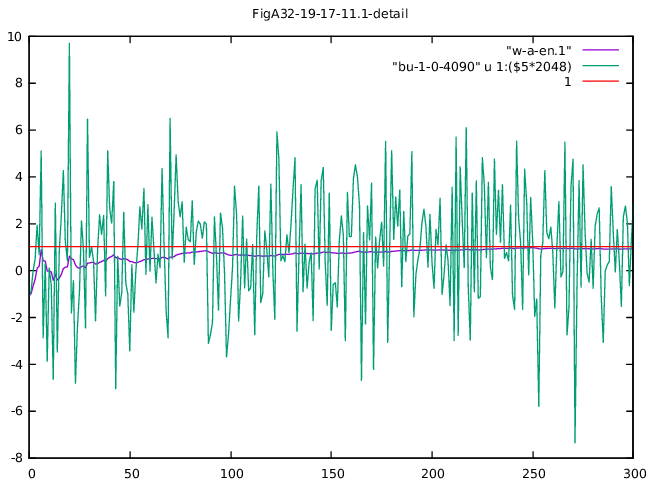}{}
%\raise 2cm\hbox{\kern7.5cm\tiny FigA32-19-17-11.1-detail}
%\vglue-20mm

\0{\it \small Fig.1: The running average of the reversible friction $R
  \a(u)\equiv R\frac{2 Re(f_{-\kk_0} u_{\kk_0})\kk_0^2}{\sum_\kk
    \kk^4|u_\kk|^2}$ (blue), superposed to the {conjectured value $1$}
  (red) and to the fluctuating values $R \a(u)$ (green)reported every 20:
  reversible NS, {R=2048}, $\NN=224$ modes, integration step $h=2^{-13}$,
  Lyapunov exponent $\l\simeq1.5$, axis unit $4h$, constant (randomly
  chosen) forcing on wave $\kk_0=\pm(2,-1)$}.

Other considerations will next be presented with suggestions
of several tests,  related to the above discussion and to ideas summarized
in \cite{Ga013b}.

The Fig.1 shows large fluctuations of the variable $\a$ for the {\it
  reversible $d=2$ NS evolution}. This variable is closely related to the
divergence $\s(\uu)=-{\rm Tr\frac{\BDpr \dot \uu}{\BDpr \uu}}$ of the
Jacobian matrix (\ie the phase space contraction): which generates a very
similar graph. Since the evolution is reversible it is tempting to check
whether the {\it fluctuation relation} (FR, \cite{GC995}) is satisfied by
the probability $\p_\t(p)dp$ that $p\defi\frac1\t\int_0^\t
\frac{\s(S_t\uu)}{\s_+}dt\in dp$ with $\s_+\defi \m^{r,N}_{En}(\s)$, \ie
whether for all UV cut-off $N$ and asymptotically as $\t\to\infty$ it is:
\be \frac{\p_\t(p)}{\p_\t(-p)}=e^{p \s_+ \t+ o(\t)}, \ |p|<1\Eq{11}\ee
The FR is a consequence, for reversible evolutions, of the {\it chaotic
  hypothesis}, \cite{GC995,Ga013b}, which states that a chaotic evolution
has an attracting smooth manifold $\AA$ on which the evolution is strongly
unstable (as in an Anosov flow). But FR is a property of the phase space
contraction $\s_\AA(\uu)$ {\it measured on the surface of the attracting
  set $\AA$}.

The latter is of course very difficult to describe analytically
as $\AA$ is in general only a (often small) subset of the surface of
constant enstrophy on which the flow takes place. Except when \\
(a) $\AA$
coincides with the entire surface of constant enstrophy $D(\uu)=En$, or
\\
(b) when, by virtue of some symmetry, the $\s_\AA(\uu)$ on the surface
of constant enstrophy is proportional to the $\s(\uu)$ on the full phase
space, which is a quantity that can be computed directly from the equations
of motion.

The question suggests studying the local Lyapunov exponents,\footnote{Local
  exponents are defined in terms of the OR decomposition of the
  linearization of the flow over a time step $h$, \ie of the Jacobian
  matrix $1+h\frac{\dpr \dot\uu}{\dpr\uu}$, averaged over the evolution of
  $\uu\to S_t\uu$, \cite{BGGS980a}, or in terms of the eigenvalues of the
  symmeric part of the Jacobian, \cite{Ru982,Li984}.} which can be quite
easily measured, and the results of a simulation can be helpful to comment
on the above items (a),(b), see Fig.2.

In the case (a) the number of positive and negative Lyapunov exponents has
to be the same. In the simulations similar to the one leading to Fig.1, but
{\it with low} UV {\it regularization} $N$, this seems to be the case, at
least if the local Lyapunov exponents are studied: therefore it would be
interesting to try to check the CH in this case, assuming that the local
Lyapunov exponents reflect at least the count of the positive and negative
ones, as it appears to be for $R=2048$ and less than $\NN=224$ modes with the
exponents recorded roughly every $4\l^{-1}$ units of time ($\l^{-1}\sim$
maximal Lyapunov exponent) and averaged. The graph of the exponents is not
reproduced here.

The case (b) arises, at the same $R$, if the UV regularization is increased,
\eg at $\NN=960$ the fraction of positive local exponents is measurably
$<\frac12\NN$. This is illustrated in Fig.2 which gives the local exponents
for both the reversible and irreversible flow in an equivalence condition.

\*
\eqfig{200}{110}{}{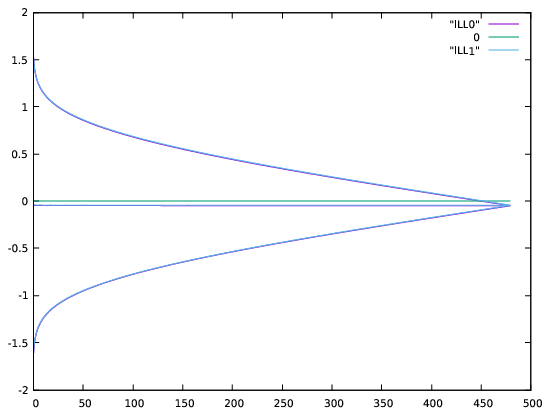}{}
%.\hglue.5cm\raise 3cm \hbox{\kern7.5cm\tiny FIGll-64-19-17-11}

\0{\it\small Fig.2: $R=2048$, $\NN=960$ modes, {\bf local} exponents ordered
  by decreasing values $\l_k,\, 0\le k < N/2$, and increasing $\l_{N-k},
  1\le k \le N/2$ and the lines $\frac12(\l_{k}+\l_{N-1-k})$, computed for
  the reversible and irreversible equations (superposed), as well as the
  line $\equiv0$ (it is hard to distinguish the latter two lines).}
\*

The first interesting feature is that the local exponents of the reversible
and irreversible cases are almost overlapping: this cannot be attributed to
the conjecture because the exponents {\it are not local observables} and
suggests that equivalence might extend, for selected observables, beyond
the local observations.

The second aspect is difficult to see on Fig.2: in it the line $0$ is
also drawn and it almost overlaps with the line $k\to
\frac12(\l_{k}+\l_{N-1-k})$. However it is {\it definitely distinct} as it
could be seen magnifying Fig.2.

The positive and negative Lyapunov exponents appear ``paired'' to a line
which is not horizontal, but close it. 

\vglue1cm
\eqfig{200}{90}{}{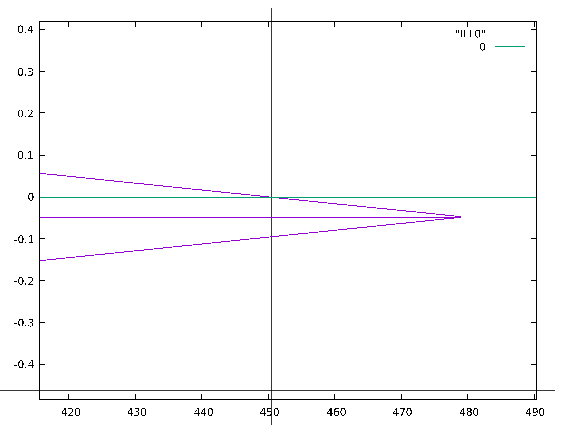}{}
%.\hglue.5cm\raise 3cm \hbox{\kern7.5cm\tiny FIGll-detail64-19-17-11}

%\kern-2.5cm
\0{\it\small Fig.3: Detail of Fig.2 showing the
  irreversible exponents (only) and the line $\equiv0$ which illustrates
  the {dimensional loss} of $\simeq\frac{450}{480}$.
  {$R=2048$, $\NN=960$ modes}.}

%\vfill\eject
Nevertheless under the, appealing
but daring, hypothesis that the pairs which consist of two negative
exponents pertain to the attraction to the attracting set $\AA$
proportionality between the total phase space contraction and the
contraction of the surface elements on $\AA$ can be establshed,
\cite[Sec.4.4]{Ga013b}. The proportionality factor turns out to be about
the ratio $\f$ between the number of pairs of exponents of opposite sign
and the total number of pairs (\ie the total dimension of $M_N$). In the
case of Fig.2,3 this would be $\f\simeq\frac{450}{480}$ and, in testing the FR,
the $\s_+$ should be replaced by $\f \s_+$: which could be observed, in
principle. However the graph draws local exponents and to find $\f$ it
would be necessary to use the Lyapunov exponents: hence the above value is,
st best, an indication of dimensional loss.
%\vfill\eject

Finally the following is a plot in a low number of modes of
the local observable $({\rm Re}\, u_{1,1})^4/{\media{{\rm Re}\, u_{1,1})^4}}$

\eqfig{160}{118}{}{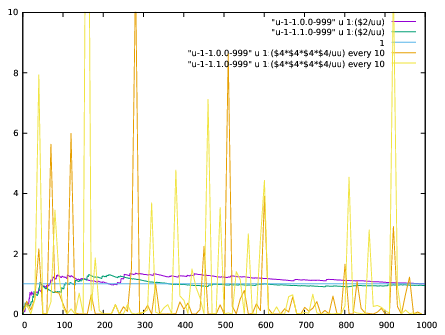}{}
%.\hglue.5cm\raise 3cm \hbox{\kern7.5cm\tiny FIG16-u4-15-13-11.01}

\def\Re{Re} \0{\it\small Fig.4: Graphs of the local observable
  $x^4/\media{x^4}$ with $x= {\rm Re}\,{u_{1,1}}$, irreversible (yellow)
  and running averages of $x^4/\media{x^4}$ superposing {reversible and
    irreversible} evolutions (blue and red) and the line $1$ (reported
  every $20$), both graphs converge to the conjectured $1$ with the
  irreversible running average converging more slowly.  $R=2048$ and
  $\NN=48$ modes ($N=3$) integration step $h=2^{-13}$ data saved every $4h$
  but reported in the graph every $20$.  Check of agreement with the
  conjectured value $1$ (straight line). }

The idea behind the conjecture is that it is due to the chaotic motion:
however the NS equations are derived from the microscopic motions which
undoubtedly are chaotic: therefore the conjecture is formulated for {\it
  all} Reynolds' number $R$. In particular at small $R$, where the motion
may be laminar (\ie periodic) or several attractors may coexist, the
equivalence will mean that the stationary states of the
reversible and irreversible equations can be put in one-to-one
correspondence in which local observables have equal average valus in the
UV limit; in close analogy with phases coexistence in statistical
mechanics.

Hence equivalence may not hold for equations which are not derived from
microscopic motions, like the fluid equations with linear viscosity
$-\n\uu_\kk$ or NS equations with {\it fixed} UV cut-off or other models,
\cite{Ga013b}, for which reversible equations have been considered. In the
latter cases some equivalence might still remain: which should hold quite
generally if at {\it fixed equations} the limit $R\to\infty$ is considered,
\cite{BCDGL018}.

{\bibliographystyle{unsrt}
\tiny
%\bibliography{0Bib}

\begin{thebibliography}{14}
\expandafter\ifx\csname natexlab\endcsname\relax\def\natexlab#1{#1}\fi
\expandafter\ifx\csname bibnamefont\endcsname\relax
  \def\bibnamefont#1{#1}\fi
\expandafter\ifx\csname bibfnamefont\endcsname\relax
  \def\bibfnamefont#1{#1}\fi
\expandafter\ifx\csname citenamefont\endcsname\relax
  \def\citenamefont#1{#1}\fi
\expandafter\ifx\csname url\endcsname\relax
  \def\url#1{\texttt{#1}}\fi
\expandafter\ifx\csname urlprefix\endcsname\relax\def\urlprefix{URL }\fi
\providecommand{\bibinfo}[2]{#2}
\providecommand{\eprint}[2][]{\url{#2}}

\bibitem[{\citenamefont{Ruelle}(1989)}]{Ru989}
\bibinfo{author}{\bibfnamefont{D.}~\bibnamefont{Ruelle}},
  \emph{\bibinfo{title}{Chaotic motions and strange attractors}}
  (\bibinfo{publisher}{Accademia Nazionale dei Lincei, Cambridge University
  Press}, \bibinfo{address}{Cambridge}, \bibinfo{year}{1989}).

\bibitem[{\citenamefont{Ruelle}(2012)}]{Ru012}
\bibinfo{author}{\bibfnamefont{D.}~\bibnamefont{Ruelle}},
  \bibinfo{journal}{Proceedings of the National Academy of Science}
  \textbf{\bibinfo{volume}{109}}, \bibinfo{pages}{20344}
  (\bibinfo{year}{2012}).

\bibitem[{\citenamefont{Ruelle}(2014)}]{Ru014}
\bibinfo{author}{\bibfnamefont{D.}~\bibnamefont{Ruelle}},
  \bibinfo{journal}{{Journal of Statistical Physics}}
  \textbf{\bibinfo{volume}{157}}, \bibinfo{pages}{205} (\bibinfo{year}{2014}).

\bibitem[{\citenamefont{She and Jackson}(1993)}]{SJ993}
\bibinfo{author}{\bibfnamefont{Z.}~\bibnamefont{She}} \bibnamefont{and}
  \bibinfo{author}{\bibfnamefont{E.}~\bibnamefont{Jackson}},
  \bibinfo{journal}{Physical Review Letters} \textbf{\bibinfo{volume}{70}},
  \bibinfo{pages}{1255} (\bibinfo{year}{1993}).

\bibitem[{\citenamefont{Gallavotti}(1997)}]{Ga997b}
\bibinfo{author}{\bibfnamefont{G.}~\bibnamefont{Gallavotti}},
  \bibinfo{journal}{Physica D} \textbf{\bibinfo{volume}{105}},
  \bibinfo{pages}{163} (\bibinfo{year}{1997}).

\bibitem[{\citenamefont{Gallavotti}(2018)}]{Ga018}
\bibinfo{author}{\bibfnamefont{G.}~\bibnamefont{Gallavotti}},
  \bibinfo{journal}{European Physics Journal Special Topics}
  \textbf{\bibinfo{volume}{227}}, \bibinfo{pages}{217} (\bibinfo{year}{2018}).

\bibitem[{\citenamefont{Ruelle}(1969, 1974)}]{Ru969}
\bibinfo{author}{\bibfnamefont{D.}~\bibnamefont{Ruelle}},
  \emph{\bibinfo{title}{Statistical Mechanics}} (\bibinfo{publisher}{Benjamin},
  \bibinfo{address}{New York}, \bibinfo{year}{1969, 1974}).

\bibitem[{\citenamefont{Gallavotti}(2005)}]{Ga002}
\bibinfo{author}{\bibfnamefont{G.}~\bibnamefont{Gallavotti}},
  \emph{\bibinfo{title}{Foundations of Fluid Dynamics}}
  (\bibinfo{publisher}{(second printing) Sprin\-ger Verlag},
  \bibinfo{address}{Berlin}, \bibinfo{year}{2005}).

\bibitem[{\citenamefont{Gallavotti}(2014)}]{Ga013b}
\bibinfo{author}{\bibfnamefont{G.}~\bibnamefont{Gallavotti}},
  \emph{\bibinfo{title}{Nonequilibrium and irreversibility}}, Theoretical and
  Mathematical Physics (\bibinfo{publisher}{Springer-Verlag and
  http://ipparco.roma1.infn.it \& ar{X}iv 1311.6448},
  \bibinfo{address}{Heidelberg}, \bibinfo{year}{2014}).

\bibitem[{\citenamefont{Gallavotti and Cohen}(1995)}]{GC995}
\bibinfo{author}{\bibfnamefont{G.}~\bibnamefont{Gallavotti}} \bibnamefont{and}
  \bibinfo{author}{\bibfnamefont{D.}~\bibnamefont{Cohen}},
  \bibinfo{journal}{Physical Review Letters} \textbf{\bibinfo{volume}{74}},
  \bibinfo{pages}{2694} (\bibinfo{year}{1995}).

\bibitem[{\citenamefont{Benettin et~al.}(1980)\citenamefont{Benettin, Galgani,
  Giorgilli, and Strelcyn}}]{BGGS980a}
\bibinfo{author}{\bibfnamefont{G.}~\bibnamefont{Benettin}},
  \bibinfo{author}{\bibfnamefont{L.}~\bibnamefont{Galgani}},
  \bibinfo{author}{\bibfnamefont{A.}~\bibnamefont{Giorgilli}},
  \bibnamefont{and} \bibinfo{author}{\bibfnamefont{J.}~\bibnamefont{Strelcyn}},
  \bibinfo{journal}{Meccanica} \textbf{\bibinfo{volume}{15}},
  \bibinfo{pages}{9} (\bibinfo{year}{1980}), ISSN \bibinfo{issn}{1572-9648}.

\bibitem[{\citenamefont{Ruelle}(1982)}]{Ru982}
\bibinfo{author}{\bibfnamefont{D.}~\bibnamefont{Ruelle}},
  \bibinfo{journal}{Communications in Mathematical Physics}
  \textbf{\bibinfo{volume}{87}}, \bibinfo{pages}{287} (\bibinfo{year}{1982}).

\bibitem[{\citenamefont{Lieb}(1984)}]{Li984}
\bibinfo{author}{\bibfnamefont{E.}~\bibnamefont{Lieb}},
  \bibinfo{journal}{Communications in Mathematical Physics}
  \textbf{\bibinfo{volume}{92}}, \bibinfo{pages}{473} (\bibinfo{year}{1984}).

\bibitem[{\citenamefont{Biferale et~al.}(2018)\citenamefont{Biferale, Cencini,
  DePietro, Gallavotti, and Lucarini}}]{BCDGL018}
\bibinfo{author}{\bibfnamefont{L.}~\bibnamefont{Biferale}},
  \bibinfo{author}{\bibfnamefont{M.}~\bibnamefont{Cencini}},
  \bibinfo{author}{\bibfnamefont{M.}~\bibnamefont{DePietro}},
  \bibinfo{author}{\bibfnamefont{G.}~\bibnamefont{Gallavotti}},
  \bibnamefont{and} \bibinfo{author}{\bibfnamefont{V.}~\bibnamefont{Lucarini}},
  \bibinfo{journal}{Physical Review E} \textbf{\bibinfo{volume}{98}},
  \bibinfo{pages}{012201} (\bibinfo{year}{2018}).

\end{thebibliography}
%\begin{thebibliography}{10}

}
\vskip3mm

\0Also:
\alert{\alertb{http://ipparco.roma1.infn.it}}\\
{Acknowledgement: I am indebted for computer support from 
  Roma 2, INFN and NYU; I am grateful to a referee for remarking several
  typos.}\\
e-mail: {\tt giovanni.gallavotti@roma1.infn.it}

\end{document}